\documentclass[preprint,amsmath]{revtex4}

\usepackage{amscd,amssymb,latexsym}
\usepackage{psfrag}
\usepackage{graphicx}

\newcommand{\ms}{\operatorname{\mathcal{S}}}
\newcommand{\br}{\operatorname{\Bbb{R}}}

\newcommand{\bv}{\operatorname{\Bbb{V}}}

\newcommand{\va}{\operatorname{\vartheta}}

\newcommand{\lo}{\longrightarrow}
\newcommand{\lom}{\longmapsto}
\newcommand{\Lo}{\Longrightarrow}
\newcommand{\la}{\langle}
\newcommand{\ra}{\rangle}

\newcommand{\ti}{\operatorname{\times}}

\begin{document}

\title{Rank change in Poisson dynamical systems}

\author{Vivek Narayanan\footnote{email: vxnsps@rit.edu \\ current address:  Physics Department, Rochester Institute of Technology,  Rochester, NY 14623-5603.}}

\author{Philip J.~Morrison\footnote{email: morrison@physics.utexas.edu}}
\affiliation{Institute for Fusion Studies and Department of Physics, 
The University of Texas at Austin, 1 University Station C1600 Austin, TX 78712-0264, USA}
 

\begin{abstract}
It is shown in this paper how a connection may be made between the symmetry generators of the Hamiltonian (or potential) invariant under a symmetry group $G$, and the subcasimirs that come about when the rank of the Poisson structure of a dynamical system drops by an even integer. This {\em kinematics-dynamics} connection is made by using the algebraic geometry of the orbit space in the vicinity of rank change, and the extra null eigenvectors of the mass matrix (Hessian with respect to symmetry generators) of the Hamiltonian (or potential). Some physical interpretations of this point of view include a control-theoretic prescription to study stability on various symplectic leaves of the Poisson structure. Methods of Invariant Theory are utilized to provide parametrization for the leaves of a Poisson dynamical system for the case where a compact Lie group acts properly on the phase space, which is assumed to be modeled by Poisson geometry. 
\end{abstract}

\maketitle
%
%
\section{Introduction}\label{intro}
 
A Poisson dynamical system is one in which a Poisson bracket between functions governs the dynamics. It can be shown that the Poisson bracket, or the rank two contravariant tensor (called Poisson structure) it defines, assumes a noncanonical form where its rank is locally constant~\cite{wei84}. Points on the phase space manifold where the rank is full are called regular, and those where the rank is less than full, singular. In many situations of interest, the change in rank governs physical properties like the presence of extra equilibria, or the stability of existing equilibria~\cite{pjmkandrup}. The characterization of singular points of the Poisson manifold (which models the dynamical system) is therefore of some interest. In what follows, we propose one such characterization for the case of a Poisson manifold that is acted on by a continuous symmetry group $G$ --- in other words, a regular $G$-Poisson manifold.  
 
The techniques we use are those that were exploited in a different context by workers in the algebraic geometry of spontaneous symmetry breaking. In such studies~\cite{abudsartori1}, it was shown how the parametrization of lower dimensional strata (comprising the so-called {\em thin} orbits) in orbit space could be carried out using the information that comes from the mass matrix (the Hessian) of the potential. We conjecture here that these extra data, that correspond to the partial restoration of symmetry in the vicinity of thin orbits, are precisely the embedding data needed to define the subcasimir surfaces. 
 
Recall that Casimirs are functions that commute with all functions, forming the center of the Poisson bracket algebra. Their gradients are null eigenvectors of the matrix representing the Poisson tensor in any chosen basis; integrating these null eigenvectors gives (up to a constant) the Casimirs for the problem. The level sets of these Casimirs would locally carve out symplectic leaves of even dimension equal to the rank of the Poisson tensor. When rank changes occur, these leaves drop in dimension by an even integer: extra Casimirs, called {\em subcasimirs}, arise and the new symplectic leaves of reduced dimension are defined by the intersection of level sets of both the Casimirs and the subcasimirs. 

Often, stability issues can be addressed by considering the Casimirs that occur on a given leaf~\cite{kruskaloberman}. Physical systems tend to equilibrate towards states of greater symmetry, which occur on dynamical leaves of greater codimension. Thus, singular leaves become relevant as arenas where actual stability issues of equilibria must be addressed. For this purpose, in addition to the Casimirs, the subcasimirs have also to be considered. The recipe we propose for thinking about the subcasimirs suggests an alternative path to studying the stability problem: namely, through the examination of stability of equilibria through the mass matrix of symmetry breaking. Said another way, we can approach dynamics through kinematics, and vice versa, by availing of the symmetry in the problem.  
 
Fuller details of many of the ideas and constructs to follow can be found in~\cite{selfthesis}. In Section~\ref{poissongeom}, we recall the basic notions from Poisson geometry. In Section~\ref{orbitspace}, Lie group actions and their orbit space structures are summarized. Symmetry breaking is introduced in Section~\ref{symmbrkng}, where we propose the characterization for subcasimirs. Section~\ref{examples} considers some examples of rank change in finite and infinite dimensions, to provide context for the proposed identification. Section~\ref{conclusion} concludes with a brief summary of the results, and directions for further investigation. 
%
%
\section{Preliminaries from Poisson Geometry}\label{poissongeom}

A Poisson manifold $P^{m}$ is an $m$-dimensional differentiable manifold with an extra structure called the {\em Poisson structure}. A Poisson structure is specified by a Lie algebra structure on the space of smooth functions $C^{\infty}(P)$, by means of an antisymmetric algebra operation called the Poisson bracket~\cite{vaisman1, weiparis}
\begin{equation*} 
\{\cdot,\cdot\}: C^{\infty}(P) \times C^{\infty}(P) \longrightarrow \Bbb{R}. 
\end{equation*}
It satisfies two properties, aside from antisymmetry,
\begin{equation}\label{jac1}
\{f, gh\} = \{f, g\}h + g\{f, h\}
\end{equation}
\begin{equation}\label{jac2}
\{f, \{g, h\}\} + \{g, \{h, f\}\} + \{h, \{f, g\}\} = 0,
\end{equation} 
which are called Leibniz's rule and Jacobi identity, respectively. 
 
In a coordinate neighborhood $\{x^{i}\}$, and for $f, g \in C^{\infty}(P)$ the above properties imply the following form for the bracket (summation over repeated indices):
\begin{equation*}
\{f, g\} =  \{x^{i}, x^{j} \}\frac{{\partial{f}}}{{\partial{x^{i}}}}\frac{{\partial{g}}}{{\partial{x^{j}}}}. 
\end{equation*}     
Setting
\begin{equation*}
\{x^{i}, x^{j} \} := J^{ij},
\end{equation*}
we see that the Poisson bracket of coordinate functions completely determines the cosymplectic form, 
\begin{equation}\label{poissontensor}
J(x) = J^{ij}(x) \frac{{\partial}}{{\partial{x^{i}}}} \bigwedge \frac{{\partial}}{{\partial{x^{j}}}}.
\end{equation}
We shall call it `Poisson tensor' or `Poisson structure' in what follows.  

\subsection{Constant rank case:}\label{regular}

If $\mbox{dim}(P) = m$, and the rank of the Poisson structure is $l$, then Darboux showed that firstly $l = 2n$, and moreover local coordinates $(q_{1}, \ldots , q_{n}, p_{1}, \ldots , p_{n}, c_{1}, \ldots , c_{k})$ exist about any point $x_{0} \in P$ such that the Poisson structure has the {\em noncanonical form} 
\begin{equation}\label{liedarboux} 
(J^{ij})|_{x_{0}} = \left( \begin{array}{ccc}
\bf{0} & \bf{1} & 0\\
\bf{-1} & \bf{0} & 0\\
0 & 0 & \bf{0_{k}}
\end{array} \right), 
\end{equation} 
which would be a canonical form except for the $\bf{0_{k}}$. 
Here, $m = 2n + k$, and zeroes appearing in the matrix have the appropriate dimensions. 

In terms of the Poisson brackets, we have the following {\em commutation relations}:
\begin{equation*} 
\{q_{i}, p_{j}\} = \delta_{ij}, \  \ \{q_{i}, q_{j}\} = \{ p_{i}, p_{j}\} = 0,
\end{equation*} 
and the $c_{k}$s commute with everything.   
 
Since $\mbox{det}(J^{ij}) = 0$, this is not the inverse of any symplectic structure. However, by the Frobenius theorem, it is possible to find {\em leaves} of a regular foliation generated by a system of vector fields of rank $2n$, which are symplectic with the symplectic form coming from (roughly speaking) the part of the matrix $(J^{ij})$ that is an invertible $2n$ by $2n$ submatrix. The remaining $k$ dimensions are transversal to the leaves and comprise the {\em transverse} structure to the foliation. For a regular foliation, the transverse structure is trivial, and is simply generated by the {\em complementary} dimension subspace of 1-form (Pfaffian) fields of rank $k$, from amongst the space of all 1-form fields. 

The leaves are regular submanifolds of $P$ that are locally just level sets of the $k$ functions in a neighborhood with generalized Darboux coordinates, $\{c_{k} = const.\}$. These are called {\em Casimirs} (or Casimir invariants), and play an important role in both the physics and geometry of Poisson dynamical systems. The Casimirs form the center of the Poisson function algebra, that is, $\{c_{k} \in C^{\infty}(P)| \{c_{k}, f\} = 0\ \ \forall f \in C^{\infty}(P)\}$. Geometrically, as indicated above, the differentials, $\{dc_{k}\}$ generate the transverse foliation of complementary dimension to the symplectic leaves. Physically, these are precisely the leaves on which dynamics, specified by a Hamiltonian, would be constrained to occur. We emphasize here that the definition of Casimirs does not call into account the Hamiltonian for the problem. The conservation of Casimirs is purely kinematical, and happens for {\em any} choice of a Hamiltonian. 

\subsection{Nonconstant rank case:}\label{singular}

If the rank of the Poisson structure is not constant as $x \in P$ varies, then the distribution of vector fields generated by the invertible part of $J^{ij}$ is integrable in the generalized sense, with leaves of foliation of varying (even) dimension~\cite{sussmann, stefan, liebmarle}.

The local structure of a Poisson manifold in the neighborhood of a point $x_{0} \in P$ was studied by Weinstein~\cite{wei84, wei98, weicontemp}. Essentially, the Poisson structure decomposes into a symplectic part and a singular part for which various forms can be postulated. The one that shall be used in this paper is the `linear structure', which entails zero rank at the origin. The {\em generic} points are those for which the symplectic part of the Poisson structure has maximum dimension. Points other than generic are called {\em singular} --- at these points, the symplectic part falls in dimension, and the singular part is one that incorporates the coordinates that have disappeared. 
 
In terms of the structure matrix, we have in a neighborhood of $x_{0}$,
\begin{equation} 
(J^{ij})|_{x_{0}} = \left( \begin{array}{ccc}
\bf{0} & \bf{1} & 0\\
\bf{-1} & \bf{0} & 0\\
0 & 0 & \pi^{ij}(y)
\end{array} \right) 
\end{equation}  
where $(q_{1}, \ldots ,q_{n}, p_{1}, \ldots , p_{n}, y_{1}, \ldots , y_{k})$ are coordinates of $P^{2n+k}$ in the above neighborhood, and 
\begin{equation*} 
\pi^{ij}(y) = \{y^{i}, y^{j}\}, \    \ \pi^{ij}(0) = 0
\end{equation*} 
is the transverse Poisson structure. In tensor notation, we have:
\begin{equation*} 
J(x)|_{x_{0}} = \sum^{n}_{i=1} \frac{{\partial}}{{\partial{q_{i}}}} \bigwedge \frac{{\partial}}{{\partial{p_{i}}}} + \sum^{k}_{j,l=1} \pi^{jl}(y)  \frac{{\partial}}{{\partial{y_{j}}}} \bigwedge \frac{{\partial}}{{\partial{y_{l}}}}.
\end{equation*}
 
In the neighborhood of any point of $x_{0} \in P$, the Poisson manifold $P^{m}$ is a product of a symplectic part and a degenerate Poisson part (in local coordinates, the expressions given above). The regular part of $P$ is an open dense subset of $P$, and here the symplectic factor arranges in leaves of locally constant dimension. The degenerate factor or transverse structure is defined up to its {\em isomorphism class}, which is the same for all points along the leaf.

By `linear structure' is meant that in a neighborhood of the point where rank is zero, the Poisson structure has a Taylor series expansion of the form 
\begin{equation*} 
J(x)|_{0} = \sum^{m}_{i,j,l=1} C^{ij}_{l}x^{l} \frac{{\partial}}{{\partial{x^{i}}}} \bigwedge \frac{{\partial}}{{\partial{x^{j}}}} + o(x^{2})
\end{equation*}
where $C^{ij}_{l}$ are structure constants of a Lie algebra $\frak{g}$ of dimension $m$ and rank $k$. 
 
If $(\mu^{1}, \ldots , \mu^{m})$ are coordinates of $\frak{g^{*}}$, $(\frac{\delta}{\delta \mu^{1}}, \ldots , \frac{\delta}{\delta \mu^{m}})$ the corresponding coordinate basis of $\frak{g}$, and $\la\mbox{ },\mbox{ }\ra$ the $\frak{g}-\frak{g^{*}}$ pairing, then for any pair of functions $f,g \in C^{\infty}(\frak{g^{*}})$, the {\em Lie--Poisson bracket} is defined by  
\begin{equation}\label{liepoissonbkt}
\{f, g\}(\mu) = \left\la\mu, \left[ \frac{\delta f}{\delta \mu}, \frac{\delta g}{\delta \mu} \right] \right\ra = C^{ij}_{l}\mu^{l}\frac{\delta f}{\delta \mu^{i}}\frac{\delta g}{\delta \mu^{j}}, 
\end{equation} 
where [ , ] is the Lie bracket on $\frak{g}$, and $C^{ij}_{k}$ the structure constants of $\frak{g}$. 
 
Thus, linearization happens when the degenerate Poisson structure at the point of total degeneracy is isomorphic to a {\em Lie--Poisson structure}. 

\subsection{An example}\label{rigidbod}
 
To this simple example we shall return later, in the context of orbit geometry. 

In 3 dimensions, a rigid body is pictured as rotating freely in space about its center of mass, under no external force. The symmetry group is $SO(3)$, acting on configuration space of three positions. The phase space group action is pictured as the coadjoint action~\cite{arnoldmath} of $SO(3)$ on the space of angular momenta, $\mu^{i} \in \frak{so(3)^{*}}$, for $i = 1, 2, 3$. 

We use the Lie--Poisson structure $J$ suggested by the structure constants $\epsilon_{ijk}$ of the Lie algebra of $SO(3)$. For two functions of the angular momenta, $f, g \in C^{\infty}(\frak{g}^{*})$, it is given by 
\begin{equation*}
\{ f, g \}(\mu) := \epsilon_{ijk} \mu^{k} \frac{{\partial}f}{{\partial}\mu^{i}} \frac{{\partial}g}{{\partial}\mu^{j}} \equiv J(df, dg),
\end{equation*}
a construct that is quite {\em independent} of the actual form of the Hamiltonian.

  The norm of the angular momentum (norm evaluated with respect to the inner product of the vector space $\frak{g}^{*}$) is a conserved quantity---
\begin{equation}\label{3dcasimir}
C = \frac{1}{2} \sum_{i=1}^{3} (\mu^{i})^{2};  
\end{equation}
from the properties of a Poisson bracket~\eqref{jac1} and~\eqref{jac2}, it follows that 
\begin{equation*}
\{ C, f \} \equiv 0
\end{equation*}
for {\em any} function $f \in C^{\infty}(\frak{g}^{*})$. Thus, Casimirs are {\em kinematical} constants of the Poisson structure itself, without any reference to the chosen Hamiltonian (dynamics). This fact also shows up in Section~\ref{symmbrkng}, where for the $G$-invariant potential, any reasonable choice can be made of a $G$-invariant function that is twice differentiable (that is, whose Hessian is defined). 
%
\section{Orbit Space Considerations}\label{orbitspace}

Consider a Lie group $G$ acting on a manifold $M$. The usual Whitney embedding theorem allows us to consider Euclidean spaces $E$ of high enough dimension instead of $M$, with the canonical Euclidean differentiable structure. A similar embedding is possible for an arbitrary action of a Lie group on a manifold. In the first instance, the (generally nonlinear) Lie group action can be replaced by a linear group action; in the second, the linear group action on $M$ can be extended to a linear action on appropriate $E$. The {\em equivariant} Whitney embedding theorem (see, for instance,~\cite{chossatbifurc}) says this can always be done. Henceforth, we restrict ourselves to linear Lie group (or, matrix group) actions on vector spaces, $G$ on $\bv^{n}$. We shall denote points of $\bv$ by $\phi \in \bv$, to be consistent with the application to symmetry breaking in Section~\ref{symmbrkng}. 
 
Further, if $G$ is compact, this linear action could be made, using the Haar measure for $G$, into an orthogonal action.\footnote{We do {\em not} need compactness to make the examples work, where the groups we consider are non-compact. For instance, the coadjoint orbits of $\frak{so(3)^*}$ and $\frak{so(2,1)^*}$ are both two-dimensional, and are defined as level sets of one Casimir each; moreover, the one Casimir is related to the other by a simple signature change of the underlying metric.} We assume this is the case for the orthogonal decompositions considered below. By this, certain directions are isolated as those that characterize a decrease in orbit dimension. 

\subsection{Orbit types}\label{types}

In what follows, basic notions in Lie group actions and standard constructions relating to {\em proper actions} will be assumed (tubular neighborhoods, slices, etc.). Good treatments are available in ~\cite{bredon, duikolk, cushmanbates}. Let $G_{\phi}$ be the {\em isotropy} subgroup (or stabilizer), that is, the set of all $g \in G$ that leaves $\phi \in \bv$ fixed. Then, $G/G_{\phi} \subseteq G$ is isomorphic to the orbit through $\phi$, $\Omega(\phi) = G \cdot \phi \subset \bv$. Any two points on the same orbit have stabilizers that are related by conjugacy through an element $g \in G$. The conjugacy class of stabilizers of a given orbit is said to constitute its {\em type}, denoted by $(G_{\phi})$. So, in summary, the orbits are isomorphic to homogeneous spaces of the form $G/H$, with $H \subseteq G$, and are of type $(H)$.  

Collect together the set of all points of $\bv$ of orbit type $(H)$, where $H \subseteq G$ is a compact subgroup. Denote this set by $\Sigma_{(H)}$.  As $H$ varies over the closed subgroups of $G$, the orbit types $(H)$ partition $\bv$ into an orbit type decomposition. 

Clearly, orbit types can vary only {\em across} the orbits, changing as the foliation goes from regular leaves to singular leaves, while remaining constant along an orbit. Also, smaller orbit types are associated to fatter orbits, which is to say, the generic orbits have the least residual symmetry. Singular orbits have smaller dimension and hence, larger isotropy subgroups that leave their points invariant, and hence larger orbit types. For the case of a Poisson manifold, it can be shown that the orbit type of a generic leaf of a foliation is always {\em Abelian}, while that of the non-generic leaf is nonabelian (Duflo \& Vergne's Theorem~\cite{encyclmath4}).

For the example of the rigid body in Section~\ref{rigidbod}, at generic points the orbits of the {\em coadjoint action} of $SO(3)$ on $\frak{so(3)^{*}}$ have residual symmetry $SO(2)$, and those of two distinct points on the same orbit (a 2-sphere) are related by a conjugate rotation, as is easily seen. The isotropy subgroup, $SO(2)$ is compact, Abelian, and its conjugacy class defines a orbit type for the union of all 2-spheres of non-zero radii. The only closed (nontrivial) subgroup of $SO(3)$ that is not conjugate to $SO(2)$ is $SO(3)$ itself, and this is the isotropy of the origin where $ \mu^{i} = 0, \forall i = 1, \ldots, 3$. Thus $\bv$, which in this case is $\frak{so(3)}^{*}$, is a disjoint union of the two orbit types, the origin of type $(SO(3))$ and everything else of type $(SO(2))$.   

Now, we mod out the space $\bv$ by the $G$-action, to get an {\em orbit space}. The reduced space $\bv/G$ is a projection of $\bv$, where each orbit in $\bv$ gets projected into a point on $\bv/G$, via the canonical projection $\Pi : \bv \longrightarrow \bv/G$. 

The decomposition of $\bv$ into orbit types is formally called a {\em stratification}. These are a bit more general than foliations, in that the strata are not required to be generated by differentiable (singular) distributions. The orbit types classify the strata in both $\bv$ and $\bv/G$ into three kinds:

\begin{enumerate}
 
\item The {\em principal stratum} $\Sigma_{P}$ of maximum orbit type $(H)$, (so that $H$ is conjugate to the isotropy subgroups for the principal orbits in the stratum), is an open dense subset of $\bv$. The image via the orbit map $\Pi: \bv \lo \bv/G$ of the principal stratum, denoted by $\hat{\Sigma}_{P}$, is connected in $\bv/G$. If $\Omega_{P}$ is a principal orbit, and $\Omega$ any other orbit, then $\exists$ a $G$-equivariant map $\Omega_{P} \lo \Omega$. If $\Omega_{P} \simeq G/H$ and $\Omega \simeq G/K$, then $K \supset H$. Also, $\pi: \Omega_{P} \lo \Omega$ is a (principal) bundle (in $\bv$), or, equivalently, $\hat{\pi}: G/H \lo G/K$ is a principal bundle (in $G$), with fiber $K/H$. 

\item If dim~$(K/H) > 0$ (or dim~$\Omega_{P} > \mbox{dim~}\Omega$), then $\Omega$ is called a {\em singular} orbit, denoted by $\Omega_{S}$, and the stratum to which it belongs is called a singular stratum $\Sigma_{S}$. The corresponding image in the orbit space is not necessarily connected, and it is denoted by $\hat{\Sigma}_{S}$.  
 
\item If dim~$(K/H) = 0$ (so that dim~$\Omega_{P} = \mbox{dim~}\Omega$) but $\Omega_{P} \lo \Omega$ is a nontrivial covering map (so that $K/H$ is finite and nontrivial), then $\Omega$ is called an exceptional orbit, the stratum to which it belongs is denoted $\Sigma_{E}$, and its image in the orbit space $\hat{\Sigma}_{E}$. 
 
\end{enumerate}
 
\noindent It can be shown on general grounds that the disjoint union of singular (and exceptional) strata form the {\em boundary} of the principal stratum. We shall not consider exceptional strata as these are not rank changing. (They would, however, be of importance in physical situations such as bifurcations, or patterns of discrete symmetry breaking.) Further, it can be shown that the minimum and maximum orbit types are unique. 

The bundle projection from a principal to a singular orbit, called $\pi$ or $\hat{\pi}$ in the definition above, can be used to devise a map, as we do in Section~\ref{symmbrkng}, in order to parameterize the embedding of the singular strata as the boundary components of the principal stratum, using ideas from symmetry breaking, and invariant theory.                      

As an example, consider the rigid body Poisson manifold. It is composed of two strata: the principal stratum $\Sigma_{P}$ being a disjoint union of all 2-spheres indexed by their radii, and of isotropy type $(SO(2))$; and a singular stratum $\Sigma_{S}$, namely the origin, of type $(SO(3))$. The corresponding orbit space strata are the projections of orbits along the half real line (also the base of a singular fibration with fibers the orbits); they are $\hat{\Sigma}_{P} = (0, \infty)$ for the principal stratum, and $\hat{\Sigma}_{S} = \{0\}$ for the singular stratum. 
 

\subsection{Tangents and Normals}\label{tangnorm}
 
To enable invariant theory parametrization to actually characterize the desired orbits, it is useful to keep track of certain subspaces that are left invariant under the group action. The dimension of these subspaces changes when a transition is made from a regular to a singular orbit. Accordingly, the class of invariants used to describe them would correspondingly change.  

Let the orthogonal action or representation be denoted by $(G, \bv^{n} \simeq \br^{n})$. $G_{\phi} = \{g \in G| \mbox{ }g \cdot \phi = \phi \}$ is the isotropy subgroup or symmetry group of a point $\phi \in \bv$, and $g_{\phi} = \{ \xi \in \frak{g}| \mbox{ } \xi \cdot \phi = \phi \}$ is the corresponding Lie algebra. The reference for what follows is~\cite{abudsartori1}.  

The tangent to an orbit $\Omega(\phi)$ in the orbit space $\bv/G$ is generated by dim~$G - \mbox{dim~}G_{\phi}$ vectors that are the image, via the action map, of as many Lie algebra elements. From the slice construction, it follows that the normal directions to the orbit are generated by a slice that is $G_{\phi}$-invariant. Not all the points in the normal direction move under the isotropy, however. The notion of an {\em invariant normal slice} plays an important role in orbit geometry. We outline the relevant definitions and properties in steps:
\begin{enumerate}
\item The tangent to the orbit, $T_{\phi}(\Omega) = \{ \xi \cdot \phi | \mbox{ } \xi \in \frak{g} \}$. Clearly, dim~$T_{\phi} = \mbox{dim~}G - \mbox{dim~}G_{\phi}$. Also, the tangent is $G$-invariant, that is $T_{g \cdot \phi} = g \cdot T_{\phi}$.   
 
\item The normal to the orbit is specified by the orthogonal decomposition, $N_{\phi} + T_{\phi} = \br^{n}$, $\la N_{\phi}, T_{\phi} \ra = 0$, with $\la \mbox{ },\mbox{ } \ra$ the $G$-invariant inner product on $\br^{n}$. That is, $N_{\phi} = \{w \in \br^{n}| \mbox{ } \la w, \xi \cdot \phi \ra = 0 \mbox{ } \forall \xi \in \frak{g} \}$, and $ \la \phi, w \ra = \la g \cdot \phi, g \cdot w \ra.$ The latter condition, given the antisymmetry of the Lie algebra to $O(n)$, when differentiated with respect to a 1-parameter group $g(t) = \mbox{exp~}t \xi$, at $t = 0$, yields the condition $\la \phi, \xi \cdot \phi \ra = 0$, which implies that the normal space contains the point on the orbit regarded now as a position vector emanating from the origin of $\br^{n}$, $\phi \in N_{\phi}$. (The zero vector, for the same reason, also lives in $N_{\phi}$ for any $\phi \in \br^{n}$.) Like $T_{\phi}$, $N_{\phi}$ is $G$-invariant, $N_{g \cdot \phi} = g \cdot N_{\phi}$. 
 
\item The set of all points in $N_{\phi}$ that are fixed under the $G_{\phi}$ action, $N_{\phi}^{0} = \{ u \in N_{\phi}| \mbox{ } h \cdot u = u, \mbox{ } \forall h \in G_{\phi} \}$, orthogonally decomposes $N_{\phi}$ further into two subspaces, 
\begin{eqnarray}\label{normaldecomp}
N_{\phi} &  = & N_{\phi}^{0} \oplus N_{\phi}^{1},\\
\br^{n} & = & T_{\phi} \oplus N_{\phi}^{0} \oplus N_{\phi}^{1}.
\end{eqnarray}
 
\item The tangent space to a {\em stratum} $\Sigma$ is well-defined, as being the orthogonal sum of the orbit tangent space and the invariant normal space: 
\begin{equation}\label{stratatangent}
T_{\phi}(\Sigma) = T_{\phi}(\Omega) \oplus N_{\phi}^{0}.
\end{equation}
Thus $N_{\phi}^{0}$ lies in the tangent space to the stratum, and $N_{\phi}^{1}$ is normal to the stratum. It can be shown that the necessary and sufficient condition that the stratum is a principal one is that $N^{0}_{\phi_{P}}$ spans the (slice) coordinates transverse to an orbit.

\item From counting dimensions,~\eqref{stratatangent} implies $\mbox{dim~}\Sigma = \mbox{dim~}G - \mbox{dim~}G_{\phi} + \mbox{dim~}N_{\phi}^{0}$. Now, use the orbit map, $\Pi: \br^{n} \lo \br^{n}/G$, under which all of the tangent space to an orbit gets mapped to the zero vector in $\br^{n}/G$. The corresponding strata also get mapped, $\Pi: \Sigma_{(\Omega)} \lo \hat{\Sigma}_{(\Omega)}$, where $(\Omega)$ indicates the orbit type of a representative orbit of the stratum. Thus, we see that $\mbox{dim~}\hat{\Sigma}_{(\Omega)} = \mbox{dim~}N_{\phi}^{0}$~\footnote{From this follows a result in particle physics called Michel's theorem: the sufficient condition that dim~$\hat{\Sigma}_{(\Omega)} = 1$ is that the restriction of $(G_{\phi}, \br^{n})$ to $(G, \br^{n})$ contains only one singlet of $G_{\phi}$.}.     
 
\item By combining the dimensionality relations above with the definitions of the various types of strata, we have a criterion for principal orbits: {\em The necessary and sufficient condition that $\Omega$ is principal is that $N_{\phi}^{1} = \emptyset$ $\forall \phi \in \Omega$, in which event, the principal stratum to which the orbit belongs is full dimension, dim~$\Sigma_{P} = n$}. 
 
\end{enumerate}


\subsection{Invariants of Orbits}\label{invariants}

In this section, we briefly indicate how invariant theory is used to parameterize orbits and the strata for a group action on a vector space. The basic result is due to Hilbert, based on which a map can be defined, which does the bookkeeping for the orbits in an algebraic manner.  

As before, let $\bv$ be a (real) vector space, on which a compact Lie group $G$ acts linearly (hence by an orthogonal representation). A polynomial $p: \bv \lo \br$ is called $G$-invariant if $p(g \cdot \phi) = p(\phi)$ $\forall g \in G,$ $\phi \in \bv$. 
 
\begin{flushleft}
{\bf Hilbert's basis theorem}\cite{langalgebra, chossatbifurc} 
\end{flushleft}
$P^{G}$, the ring of $G$-invariant polynomials under the operations of polynomial addition and multiplication, is a {\em finitely} generated algebra under $\br$.\\ 
 
\noindent The algebra $P^{G}$ is a graded algebra, graded by the degree of the invariant polynomial. The basis guaranteed by the above theorem is called an {\em integrity} basis. There is a minimum number $q$, for a given representation $(G, \bv)$; this is called a minimal integrity basis, or Hilbert basis. We denote the generators of $P^{G}$ by $(\theta_{1}, \ldots, \theta_{q}) \in \br^{q}$. Any polynomial $p \in P^{G}$ can be written, $\phi \in \bv$, $p(\phi) = \hat{p}(\theta_{1}(\phi), \ldots, \theta_{q}(\phi))$. When a Hilbert basis is formed by algebraic independent polynomials, it is called {\em free} and the corresponding representation $(G, \br^{n})$ is called {\em cofree}. A non-cofree basis would have a certain number of nontrivial algebraic identities between the basis elements. (All cofree representations of complex semisimple Lie groups have been classified by G.~W.~Schwarz~\cite{schwarz}.) 
 
Next we define the {\em Hilbert map} (sometimes called the orbit map; but we reserve that term for $\Pi: \br^{n} \lo \br^{n}/G$) as a map that goes from the orbits to its invariant parametrization, $\frak{H}: \bv^{n} \lo \br^{q}$, $\phi \lom (\theta_{1}(\phi), \ldots, \theta_{q}(\phi))$. The image $\frak{H}(\bv)$ is a {\em semi-algebraic} variety $\ms \in \br^{q}$. Via the orbit map $\Pi$, $\br^{n}/G$ becomes immersed as a semi-algebraic variety~\footnote{An algebraic variety is a zero locus of a finite set of polynomials. A semi-algebraic variety is defined in part by inequalities~\cite{griffiths}.} $\hat{\ms} \in \br^{q}$ as well. 
 
The main theorem that allows orbits to be parametrized by invariants is the following:
\begin{flushleft}
{\bf Theorem}
\end{flushleft}
\begin{enumerate}
\item $\frak{H}$ is a proper map (namely, the inverse image of a compact set is compact also).
 
\item If $\phi \not \in \Omega(\phi')$, then  $\frak{H}(\phi) \neq \frak{H}(\phi')$; points from distinct orbits are {\em separated} by the Hilbert map. 
 
\item $\ms \simeq \hat{\ms}$, that is, there is a `commutative triangle', 
\begin{equation*}
\begin{CD}
\br^{n}@>{\frak{H}}>>\ms \in \br^{q}\\
@V{\Pi}VV   @.{\hspace{-.3 in}\parallel} \\
\br^{n}/G@>>{\frak{\hat{H}}}>\hat{\ms} \in \br^{q}  
\end{CD}
\end{equation*} 
where the map $\frak{\hat{H}}$ is a homeomorphism. 
\end{enumerate}
  
The above result for polynomials has been generalized to the {\em smooth} category by G~.W.~Schwarz. The proof essentially uses Weierstra\ss's approximation theorem. A smooth function may also be written as a function of a finite number of basis elements that generate the ring of smooth $G$-invariant functions.    

\begin{flushleft}
{\em \underline{Examples}}
\end{flushleft}
As an example, let us consider $SU(n)$ in the adjoint representation, defined by $Tr \phi = 0$. A basis for the invariants are just the Casimirs, $\theta_{1} = Tr \phi^{2}, \ldots, \theta_{n-1} = Tr \phi^{n}$. Similarly for $\frak{so(m)}$, the Lie algebra of skew-symmetric matrices, $X^{T} = -X$, for which the invariants are $S_{k} = Tr X^{k}$, $k = 1, \ldots, [n/2]=l$. The characteristic polynomial is the expression 
\begin{eqnarray*}
\Delta_{X}(\lambda) & = & |X - \lambda \boldsymbol{1}|\\
& = & \lambda^{n} - p_{1}\lambda^{n-1} - \cdots - p_{n},
\end{eqnarray*}
The various $p_{i}$ in the expansion are the $i^{th}$-symmetric function of the eigenvalues of $X$. Let the polynomial ring over the real numbers generated by these $p_{i}$ be denoted by $\br[p_{1}, \ldots, p_{n}]$. Clearly, the symmetric functions form an integrity basis for an arbitrary symmetric polynomial in $n$ variables. There is an {\em isomorphism} between $\br[p_{1}, \ldots, p_{n}]$ and $\br[S_{1}, \ldots, S_{l}]$, made explicit through a set of recursive relations known as the {\em Newton formulas}, 
\begin{eqnarray*}
S_{i} - p_{1}S_{i-1} + p_{2}S_{i-2} - \cdots + \\
+(-1)^{i-1}p_{i-1}S_{1} + (-1)^{i}p_{i} & = & 0, \\
\mbox{\hspace{20mm}}i = 1, \ldots, n\\
S_{i} -p_{1}S_{i-1} + p_{2}S_{i-2} - \cdots +  \\
+(-1)^{n}p_{n}S_{i-n} & = & 0,\\
\mbox{\hspace{20mm}}i = n+1, n+2, \cdots.
\end{eqnarray*}

It is in fact possible~\cite{boyarski1} to use these relations in order to parameterize singular orbits of the $SO(n)$ coadjoint action on the Lie coalgebra $\frak{so(n)}^{*}$, and in general, for any semisimple Lie group. The various symmetric functions $p_{1}, \ldots, p_{n-k}$ could be written as the sum over squares of Pfaffians (essentially determinants of minors obtained by deleting the complementary indices). Setting them equal to zero amounts to looking at lower rank subalgebras of $X \in \frak{so(n)}$, or on the group level, at homogeneous spaces of the form (for $n = 2m$ case as an example) $SO(2m)/(SO(2k) \ti U(1)^{m-k}$. From general considerations of orbit types (Section~\ref{tangnorm}) and symmetry breaking (dealt with below), this amounts to a {\em partial breaking} of symmetry. The subcasimirs that arise in the process are simply the coordinates of the embedding specified by the above system of equations. (The representation in this case is called {\em non-cofree} as the equations are tied up with determinantal identities.)

\section{Symmetry breaking}\label{symmbrkng}

Now, we suppose a Lie group $G$ of appropriate type acts on the Poisson manifold, so that the leaves of the Poisson manifold are contained in the orbits of the group action. In particular, we choose the group to be such that its rank (defined as the dimension of its Cartan subalgebra) coincides with the {\em corank} of the Poisson structure at its generic points. In making this ansatz, we rule out several interesting Poisson structures~\cite{marsdeninternetapp}, where the action of a group on the manifold is globally transitive (for example, the action of $SO(2)$ on a 2-torus $\Bbb{T}^{2}$, with irrational winding number). We only consider here group actions for which Casimirs define both the group orbits as well as the Poisson leaves. Such a requirement is automatically satisfied by the much investigated {\em Lie--Poisson} structures. There, a semisimple Lie group acts on its coadjoint space, and the Casimirs form the center of the coadjoint representation. But in fact, this is generically so in virtue of the Linearization conjecture of Weinstein, cf. Section~\ref{singular}. In this section, for any reasonable choice of $G$-invariant Hamiltonian (or potential), it is seen that by the process of symmetry breaking, the null eigenvectors of the mass matrix span out the directions {\em tangent} to the orbits. Methods of invariant theory are then used to characterize the various orbits. Remembering that {\em some} of these orbits describe the Poisson leaves, we make the identification of these invariants, which come from symmetry breaking, to the Casimirs and subcasimirs of Poisson geometry, and provide in the process a dynamical interpretation for a kinematical phenomenon.   

Let us suppose that symmetry is broken from the original symmetry group $G$ to a reduced symmetry group, $H$.\footnote{Symmetry Breaking is equivalent, mathematically, to the procedure of reduction, and either procedure can be viewed as principal bundle reduction~\cite{marsdenratiu, abrahammars}.} The null eigenvectors of the mass matrix, or Goldstone modes, are parametrized by the homogeneous space $G/H$. We note that the generic leaf corresponds to a maximum breaking of symmetry --- the original symmetry group is reduced to its (Abelian) Cartan
torus, denoted by $G_c$. Then, $G/G_c$ is isomorphic to a principal orbit. Partial breaking of symmetry corresponds to leaves where some of the symmetry has been {\em restored}. The stabilizers of these leaves are of higher dimension, and also non-abelian, and so the leaves, by virtue of the bundle map $\pi$ (or, $\hat{\pi}$, cf. Section~\ref{types}), are singular leaves. (Singular leaves can be formally studied as singular reductions of manifolds with symmetry, and this program is carried out in~\cite{ratiuortega, sjamaarthesis, lermanetal}, and the references cited therein.)  
 
For the case that the Poisson manifold $P = \frak{g}^{*}$, where $\frak{g}$ is a semisimple Lie algebra and  $\frak{g}^{*}$ its dual, this identification is the natural one suggested by the {\em maximal isotropy embedding} of a Lie algebra $\frak{h}$ in the larger Lie algebra $\frak{g}$, for the case that symmetry is broken from $G$ down to $H$~\cite{oraferti}. We are only concerned with patterns of symmetry breaking that entail a rank change, and assume that this can be arranged.  

We now make these observations physically relevant employing ideas from spontaneous symmetry breaking, with the added feature of incorporating free parameters, or {\em controls}. The notion of {\em patterns} in the breaking of symmetry, namely variation in the subgroups of the original symmetry group that survive when control parameters (like coupling constants, or classical masses) are varied, played an important historical role in spontaneous and gauge symmetry breaking in particle physics. Here, the hadronic potential was assumed to have a certain polynomial form (say, quartic), and the problem was to find the minima of the potential by varying with respect to certain parameters. This problem was soon realized to be mostly geometrical in character~\cite{kim, michel, abudsartori1, abudsartori2, abudsartori3, sartori4}, with orbit geometry and invariant theory playing a role in defining the possible orbital strata that the minima could occupy. We turn this development around, and use arbitrarily chosen $G$-invariant potentials and their extremal values, in order to relate kinematical data of the Poisson bracket, with dynamical information that comes from symmetry breaking by the potential.

\subsection{Bosonic symmetry breaking}\label{bosonic}

This is the most basic mechanism for the idea of symmetry breaking. The G-invariant Hamiltonian is a function of a certain number of scalar fields. The fields themselves depend on the spacetime coordinates, and the symmetry group acts on the space of fields. Spontaneous breaking of symmetry is said to occur when the symmetry of the solutions got by extremizing the potential, is less than the full symmetry group~\cite{ryder}. The material to follow is fairly standard, and our purpose is to establish notation, and reinterpret standard results in the light of orbit geometry. 

Let $\bv^{n}$ be the space of $n$ scalar fields, $\phi_{i}: M \lo \Bbb{R}$, where $M$ is some underlying differentiable manifold (not relevant for our purposes). Let $G$ be a compact Lie group that acts properly on $\bv$, and let $V(\phi)$ be a $G$-invariant scalar potential function on $\bv$. Then the condition for a {\em vacuum} is that 
\begin{equation*}
\left. \frac{{\partial}V(\phi)}{{\partial}\phi_{i}} \right|_{\phi = \phi_{(e)}} = 0, \mathbb{\hspace{10mm}} (i = 1, \ldots , n),
\end{equation*}     
so the Taylor expansion of the potential about equilibria takes the form:
\begin{equation*}
V(\phi) = V(\phi_{(e)}) + \frac{1}{2} M_{ij} \chi_{i} \chi_{j} + o(\chi^{3}).
\end{equation*}
In the above expression, the `mass matrix' $M_{ij} \geq 0$ is defined as
\begin{equation}\label{massmatrix}    
M_{ij} = \left. \frac{{\partial}^{2}V(\phi)}{{\partial} \phi_{i}{\partial} \phi_{j}} \right|_{\phi_{(e)}}, 
\end{equation}
and the shifted fields by $\chi_{i}(x) = \phi_{i}(x) - \phi_{(e)i}$. The mass matrix, being the second derivative of the potential, governs the stability of the equilibrium. (We refer the reader to~\cite{pjmelizier} for more on stability in field theory systems; we do not concern ourselves with this aspect of the theory, but rather simply use the null spectrum of $M_{ij}$ to study orbit geometry on $\bv$, which models the linear Poisson $G$-manifold. See however, the conclusions, Section~\ref{conclusion}.)   

Since the potential is $G$-invariant, it satisfies the invariance condition at the vacuum as well, so 
\begin{equation}\label{ginvofv}
V(\phi_{(e)}) = V(\Theta_{g}\phi_{(e)}) = V(\phi_{(e)}) + \frac{1}{2} M_{ij} \delta \phi_{(e)i} \delta \phi_{(e)j} + \cdots,
\end{equation}
where $\Theta_{g}: \bv \lo \bv$ is the action $\forall g \in G$.

\noindent Let $(T^{a})^{i}_{j}$ denote the representation matrices, where $a = 1, \ldots, d = \mbox{dim~}G$, and $i,j = 1, \ldots, n = \mbox{dim~}\bv$, so that $\Theta_{g} = \mbox{exp~}tT^{a} \xi_{a}$, $\{ \xi_{a} \} \in \frak{g}$, $g(t) \in G$ a 1-parameter subgroup, and set the variation $\delta \phi_{(e)}$ to be:
\begin{equation*} 
\delta \phi_{(e)i} = \left[ \left. \frac{{\partial}\Theta_{g}}{{\partial}\xi_{a}} \right|_{t=0} \phi_{(e)} \right]_{i} \delta \xi_{a}. 
\end{equation*}       
 
Now, suppose that only a subgroup $H \subset G$ leaves invariant the equilibrium point. We have two possibilities:
\begin{enumerate}
 
\item If $g \in H$, then $\delta \phi_{(e)i} \equiv 0$, and the $G$-invariance condition for $V$~\eqref{ginvofv} is identically satisfied, that is $V(\phi_{(e)}) = V(\phi_{(e)})$, since $M_{ij} \delta \phi_{(e)i} \delta \phi_{(e)j} + \cdots \equiv 0$. 
 
\item If $g \in G/H$, then $\delta \phi_{(e)i} \neq 0$, and so $G$-invariance of $V$ now would require that $M_{ij}$ have null eigenvectors:
\begin{equation*}
M_{ij}[\Theta'|_0 \phi_{(e)}]_{j} = 0.
\end{equation*}   
There are dim~$G/H = \mbox{dim~}G - \mbox{dim~}H$ null eigenvectors of the form $\{\Theta'|_0 \phi_{(e)} \}$, of eigenvalue zero, that represent massless fields, called Goldstone (or Nambu-Goldstone) bosons, and the homogeneous space $G/H$ parameterizes the space of Goldstone bosons via the action map $\Theta: G \ti \bv \lo \bv$. 
\end{enumerate}
 
Before we move on to the next subsection, a few comments are in order. The number of massive fields (those with nonzero eigenvalues to the mass matrix) is dim~$H$. If $H$ happens to be the isotropy subgroup $G_{\phi_{(e)}}$ for the equilibrium point, $\phi_{(e)}$, then these modes are transversal to the orbit of the group action, and represent the `difficulty' in getting away from the orbit, as they are massive modes, and doing so would cost energy. Suppose now the potential depended on some number of control parameters (these could be coupling constants, classical masses, charges, background field strengths, etc.), that is, we write $V = V(\phi, \gamma)$, where $\gamma$ are the controls. Extremizing the potential (supposed $G$-invariant) as above, we would arrive at equilibria that depended on the control parameters, $\phi_{(e)} = \phi_{(e)}(\gamma)$.     
   
As the controls are varied, we are led from solution to solution. Formally, this happens in an infinite dimensional space of solutions (moduli space) of the classical potential~\footnote{A generalization to the simple potential extremization we have used  would involve a full Hamiltonian or action, that may preclude a splitting into potential and kinetic terms. Then we cannot implicitly assume the kinetic terms are already frozen, or set to zero. Interesting situations like {\em Killing symmetry} directions could thus arise.}, but we shall restrict our attention to what goes on in the orbit space decomposition (cf. Section~\ref{types}) for $\bv$. For some values of the control parameters, the isotropy subgroup of the solution may change, so we are led from leaves of one dimension to leaves of other dimensions, and across orbit types. Then invariant theory (cf. Section~\ref{invariants}) shall relate the gradients of the potential and the null eigenvalues of the mass matrix. Referring all this back to the Poisson manifold we began with, the picture we seek would be complete.   

\subsection{Extrema and strata}\label{extrema}

In this section, we use the invariants provided by the gradients of the $G$-invariant potential $V(\phi, \gamma)$, with $\phi \in \bv^{n}$ and $\gamma \in Q^{s}$, where the latter notation means the control parameters are derived from some $s$-dimensional manifold of controls. (In what follows, the structure of $Q$ will play no role, although its dimension is relevant.) The orbit in $\bv$ to which the equilibrium belongs is characterized by the orbital stratum, and the equations of the stratum are carved out by these gradients. In making this relation, the geometry of the tangent space and normal space to an orbit (cf. Section~\ref{tangnorm}) shall be extensively invoked. 

Recall that the tangent space to the principal stratum comprises both the tangent space to the orbit, $T_{\phi}(\Omega) = \{ \xi \cdot \phi|\mbox{ } \xi \in \frak{g} \}$, and the $G_{\phi}$-invariant normal slice $N_{\phi}^{0}$ transversal to it. By the results stated in Section~\ref{tangnorm}, the main observation about the generators of $N_{\phi}^{0}$ is the following: the invariant normal slice is the span of $\partial \theta_{a}(\phi)$, $a = 1, \ldots, q$. The proof may be found in~\cite{abudsartori3}. Note that $q$ is the minimal number of basis elements $\theta_{i}(\phi)$, guaranteed to exist by Hilbert's theorem.
 
Furthermore, any orbital invariant on $\bv$, such as a $G$-invariant potential for symmetry breaking, can be expressed in terms of the Hilbert basis as a function in the orbit space $\bv/G$:
\begin{equation}\label{orbinv}
V(\phi) = \hat{V}(\theta_{1}(\phi), \ldots, \theta_{q}(\phi)),
\end{equation}
$\forall \phi \in \br^{n}$. It follows at once by differentiating the $G$-invariance condition for the potential that its gradients span the invariant normal slice: $V(g \cdot \phi) = V(\phi) \Lo \la \xi \cdot \phi, {\partial}V(\phi) \ra = 0$ for every one parameter subgroup $g(t) = \mbox{exp~}t \xi \in G$; by referring to Section~\ref{tangnorm}, this is seen to be the condition for an invariant  normal vector. We note that {\em any} $G$-invariant function on the orbits has the spanning property; so the orbit geometry is a lot more general than the system specifically requires. However, a physical potential (or Hamiltonian) naturally allows for controls (masses, coupling constants, various length scales, etc.), and these can be manipulated to enable wandering across orbit types.  
 
The final step in the process of using gradients to characterize strata is counting dimensions. Let $\phi_{P}$ denote a point that belongs to a principal stratum, and $\phi_{S}$ to a singular or nonprincipal stratum. We note that from the relation~\eqref{normaldecomp}, it follows that the sum of the dimensions of the orbit tangent $T_{\phi}$ and of the two normal components, $N_{\phi}^{0}$ and $N_{\phi}^{1}$ is fixed, at $n$ = dim~$\bv$. Since the singular stratum bounds the principal stratum, $\Sigma_{S} \subset \partial \Sigma_{P}$, thinner orbits have fatter normal spaces, and these are characterized by the appearance of extra gradients. These gradients will, however, not be $G_{\phi_{P}}$-invariant, that is, they get moved around in the `extra' dimensions transverse to the thinner orbit. The larger invariance group, $G_{\phi_{S}} \supset G_{\phi_{P}}$ means at once two things: a larger number of invariants (that are left invariant with respect to the larger subgroup); and a larger number of dependent relations (that is, the representation when restricted to the singular orbit is no longer cofree).

Next, we refer to the orbit space, $\bv/G$, where distinct points are separated by invariants using the Hilbert map; this map takes both the vector space strata and the orbit space strata to the corresponding semialgebraic varieties $\mathcal{S}$ and $\hat{\mathcal{S}}$ both of which live in the Euclidean space $\br^{q}$. 

Define the real, symmetric, positive semi-definite (no eigenvalues less than zero) matrix formed of the gradients of the invariant polynomials, 
\begin{eqnarray*}
R_{ab}(\phi) & = & \sum_{i=1}^{n} \partial_{i}\theta_a(\phi)\partial_{i}\theta_b(\phi)\\    
& = & \hat{R}_{ab}(\theta(\phi)),
\end{eqnarray*}
with $a, b = 1, \ldots, q$ and $\partial_i$ stands for derivative w.r.t. $\phi_{i}$.
 
\noindent The number of independent vectors in $\{\partial \theta_a \}$ is equal to rank~$\hat{R}_{ab}(\phi)$ = dim$~N_{\phi}^{0}$. The strata and their images in the orbit space are carved out by the locus of the determinants of the principal minors of the above matrix (compare the example from Section~\ref{invariants}). The gradients of these determinants in the orbit space can be shown to span the null space of the matrix $\hat{R}_{ab}$, defining the `normal' directions for the image of the stratum to which $\phi$ belongs.  
 
Introducing a potential that needs to be extremized is the same as finding the null eigenvectors of the $R$ matrix above, since $\partial_i V(\phi) = 0$ implies, by using~\eqref{orbinv} the same as $\sum_{a=1}^{q} \hat{R}_{ab}(\theta)\partial_b \hat{V}(\theta) = 0$. This is a locus of algebraic equations and inequalities in orbit space $\bv/G$ (a semialgebraic variety) with the determinants of the minors of corank the dimension of the strata $\hat{\Sigma}_{\phi}$ serving as the Lagrange multipliers. Extremizing this system of equations yields an optimal set of values for both basis invariants and the control parameters; the mass matrix~\eqref{massmatrix} may be then formed out of these.~\footnote{The number of zero eigenvalues of the mass matrix gives the corank of the corresponding orbit Lie-Poisson structure.}    

\subsection{Goldstone modes and Poisson geometry}\label{goldstones}

For a Poisson manifold $P$, the Casimirs arise as constraint surfaces that are defined by a generalized distribution associated to the Poisson structure $J$. Although the procedure for finding subcasimirs reduces to computing the null eigenvectors of the Poisson structure matrix restricted to the orbit in question, our aim here is to connect the subcasimirs with null eigenvectors of the mass matrix~\eqref{massmatrix}. With controls added, the mass matrix has the form, 
\begin{equation*}
M_{ij} = \partial_i \partial_j V(\phi,\bar{\gamma})|_{\phi = \phi_{(e)}},
\end{equation*}
$i, j= 1, \ldots, n$. Here, the controls themselves have been solved with the principal minor determinants added to the potential as constraints and extremized collectively (to yield $\bar{\gamma}$) with respect to both the invariant polynomials and the Lagrange multipliers (as many added as there are normal directions to the orbit space stratum, $\hat{\Sigma}$). We implicitly assume that any smooth variation $\delta \gamma$ in the controls changes the vacuum solution {\em perturbatively}; that is, if $\delta \gamma \lo 0$, then $\delta \phi_{(e)} \lo 0$ as well.   
 
Now, it can be shown using the above, that the mass matrix decomposes into two parts: $M^{1}$ and $M^{2}$, such that $M^{1} \cdot N_{\phi_{(e)}}^{0} = 0$, $M^{2} \cdot N_{\phi_{(e)}}^{1} = 0$, in addition to the usual equations for the scalar Goldstone modes tangent to orbit: $ M^{1} \cdot T_{\phi} = 0 = M^{2} \cdot T_{\phi}$.  

The former Goldstone bosons correspond to the directions which arise from adding the constraint conditions, and must therefore be regarded as new features that arise due to the orbital structure~\footnote{In physics, one of the components, namely the $N_{\phi}^{0}$ has an interpretation of being associated to pseudo-Goldstone modes, which do not survive radiative corrections~\cite{georgpais, weinberg}.}. In the context that we are working within, these are precisely the subcasimirs that would arise upon a process of partial {\em symmetry restoration}, where the isotropy group of the equilibrium segues into one of larger dimension (and does so in a continuous manner since it is perturbative). Let $\phi_{(e)}$ denote the generic vacuum, with isotropy $G_{\phi_{(e)}}$, and $\phi_{(e')}$ the new vacuum upon variation of the control parameters, with isotropy $G_{\phi_{(e')}}$, with $G_{\phi_{(e')}} \supset G_{\phi_{(e)}}$. Conversely, when the controls are reversed, these extra Goldstone modes from the normal directions become the Goldstone bosons for a symmetry breakdown from the larger isotropy group $G_{\phi_{(e')}}$ to the smaller one, $G_{\phi_{(e)}}$. The number of extra Goldstone modes that are needed for this is, by Goldstone's theorem, equal to the difference in the orbit dimensions of the fatter and thinner orbits, which is the dimension of the homogeneous space $G_{\phi_{(e')}}/G_{\phi_{(e)}}$. These new modes may be regarded as the embedding coordinates for the singular leaf into a regular leaf. Formally, the {\em Symmetry Breaking map} (SB) may be defined as the projection $\hat{\pi}: G/G_{\phi_e} \lo G/G_{\phi_e'}$, as defined in Section~\ref{types}, and the {\em Symmetry Restoration map} (SR) is simply the inverse of the above bundle map. 

As an illustration, to connect with the $SO(3)$ case, imagine the isotropy of the ``double well'' solution to increase from $U(1)$ at the trough (flat direction), to $SO(3)$ at the top of the hillock of the well, at the center. Clearly the new equilibrium, for any intermediate hillock shape, is an unstable one, a fact reflected in the degeneracy in the modified mass matrix. (Thus, without the controls, the origin would never become a preferred point of rest. It would, if the shape were modified so much by the controls that it became favorable to rest there.) In general, if the equilibrium is a stable (or neutrally stable) one, then the pseudo-Goldstone bosons are absent; if the latter manifest themselves, then stability is lost. It would be very interesting to see if such equilibrium analyses for the examples of Section~\ref{examples} below, using the methods of invariant theory and symmetry breaking as above, would yield a similar simple stability criterion.     
  

\section{Some examples of rank change}\label{examples}

We consider situations where rank change phenomena manifest themselves, and to which the foregoing observations apply. Details and background material are provided in~\cite{selfthesis}. We begin with the $n$-dimensional generalization of the rigid body, and end with the prototype for an infinite-dimensional context for rank-change. A word of caution is in order: while the preceding sections require that $G$ be compact, the examples below do not assume this. As observed in the first footnote to Section~\ref{orbitspace}, this does not change the counting of the Casimirs or subcasimirs, although the topology of the level sets they would define (even the number and types of connected components!) can be wildly different for the compact and the noncompact cases, as we see below. 

\subsection{n-D rigid body}\label{liepoisson}
 
The symmetry group is taken to be $SO(n)$, or as a generalization, any semisimple Lie group, $G$ (in which event the term `rigid body' ought not to be taken literally). As in the 3-dimensional case, we write the Lie-Poisson bracket~\eqref{liepoissonbkt} to be the one suggested by the structure constants of the Lie algebra $\frak{g}$; for any functions $f, g \in C^{\infty}(\frak{g}^{*})$ we set  
\begin{equation}\label{liepoisson}
\{ f, g \}(\mu) := \left \la \mu, \left[\mbox{ }\frac{{\delta}f}{{\delta \mu}},\frac{{\delta}g}{{\delta \mu}}\mbox{ } \right] \right \ra = \mu^{i} \left[ \mbox{ }\frac{{\delta}f}{{\delta \mu}},\frac{{\delta}g}{{\delta \mu}}\mbox{ }\right]_{i},
\end{equation}   
where 
$\frac{{\delta}f}{{\delta \mu}} = (\frac{{\delta}f}{{\delta \mu}})^{l}e_{(l)}$, $\frac{{\delta}g}{{\delta \mu}} = (\frac{{\delta}f}{{\delta \mu}})^{m}e_{(m)}$, where $\{e_{(i)}\}$ is a basis for $\frak{g}^{*}$, and the inner bracket $[\mbox{ }, \mbox{ }]$ stands for the Lie algebra bracket, 
\begin{equation*}
[e_{(l)}, e_{(m)}] = C_{nlm} e_{(n)}
\end{equation*}
and $C_{nlm}$ the structure constants of the Lie algebra. The Poisson tensor now becomes 
\begin{equation}
J^{lm} = C^{lm}_{n} \mu^{n}.
\end{equation}

The origin is once again a point of total degeneracy, where rank of the Poisson tensor is zero. Elsewhere, it is equal to the {\em corank} of the Lie algebra, which is also the codimension of its Cartan subalgebra. In fact, the Cartan torus is generated by the {\em duals} of the differentials of the Casimirs, which are each responsible for one $U(1) \simeq SO(2)$ factor. These various $U(1)$ factors commute, as they must for a torus, and this fact is captured in one of the orbit theorems that distinguish between regular and singular points. When rank of $J$ is less than full, the differentials of the Casimirs of the reduced dimensional orbit, which now include subcasimirs, no longer commute, and hence form a nonabelian subalgebra of $\frak{g}$. 

Elsewhere, away from the origin, the leaves have the dimensionality given by the difference between dimension $d$ of $\frak{g}$ and its rank $r$, which is always an {\em even} integer, as it also equals the number of roots (which are none other than weights of the adjoint representation). So the orbits are always even dimensional, and there is in fact a nondegenerate symplectic form (the Kirillov-Kostant form) that inverts~\eqref{liepoisson} away from the origin~\footnote{We emphasize the distinct usage of the word `rank' for {\em both} the Lie algebra and the Lie-Poisson structure on its dual: the former refers to the number of Casimirs, while the latter refers to the coadjoint orbit dimension. We continue using the commonly used term `rank' for either case, taking care to note that rank of a Lie algebra is numerically equal to the corank of the Lie-Poisson structure on its dual!}.

Let us consider a few semisimple Lie algebras. $G = SU(n)$ has dim~$G = d = n^{2} - 1$, rank~$G = r =  n - 1$, so $d - r = n(n - 1)$, which is always even. SO(n) has $d = n(n - 1)/2$ and $r = (n - 1)/2$ ($n$ odd), or $r = n/2$ $n$ even). In either event, $d - r = (n - 1)^{2}/2$ ($n$ odd), or $d - r = n^{2}/4$ ($n$ even). Similar situations obtain for other Lie algebras. Symmetry breaking and restoration via maps SB and SR (see Section~\ref{goldstones}) away from the origin occur between two leaves of non-maximal isotropy. But it must be emphasized that the totality of these phenomena involving SB and SR maps far exceeds the actual number of Poisson structures that could be put on a semisimple Lie group. With some exceptions, in most Lie-Poisson groups, the only degenerate leaf is the origin. That is, admissible Poisson structures have orbit structure that are a small subset of the possible structures that could be arranged through a $G$-invariant potential. Indeed, in some cases, an SB or SR map need not be rank-changing~\cite{oraferti}. Examples of the latter occur when partial symmetry breaking occurs with respect to product groups, so that the leaf dimension decreases while the rank remains the same if the product factors are arranged to come from the diagonal of the original group. 

\subsection{Rigid body in gravity and underwater}

Departing from the notation of the 3-D rigid body a bit, we denote by $\vec{\mu} = (\mu_{1}, \mu_{2}, \mu_{3})$, the angular momentum of the body, and by $\vec{z} = (z_{1}, z_{2}, z_{3})$, the (constant) gravitational vector from the center of mass. The configuration space is the 6-dimensional vector $(\vec{\mu}, \vec{z})$. $SO(3)$ acts according to the semidirect product bracket, so we write $W = (SO(3) \ltimes_{\va} \Bbb{R}^{3})$, where $\va$ is the Adjoint representation, and define the product rule at the group level
\begin{equation*}
(R, v) \cdot (R', v') = (RR', Rv'+v)
\end{equation*}
with $R, R' \in SO(3)$, $v, v' \in \Bbb{R}^{3}$. 
 
The dual of the Lie algebra of the semidirect product group is $\frak{w}^{*} = \frak{so(3)}^{*} \ltimes_{\va^{*}} \Bbb{R}^{3}$, where $\va^{*}$ is the coadjoint representation, with the semidirect product Poisson structure given by 
\begin{equation}\label{sdpgravitymatrix}
J(\vec{\mu}, \vec{z}) = \left( \begin{array}{ccccccc}
0 & \mu_{3} & \mu_{2} & \vdots & 0 & z_{3} & z_{2}\\
-\mu_{3} & 0 & -\mu_{1} & \vdots & -z_{3} & 0 & -z_{1}\\
-\mu_{2} & \mu_{1} & 0 & \vdots & -z_{2} & z_{1} & 0\\
\hdotsfor{7}\\
0 & z_{3} & z_{2} & \vdots & 0 & 0 & 0\\
-z_{3} & 0 & -z_{1} & \vdots & 0 & 0 & 0\\
-z_{2} & z_{1} & 0 & \vdots & 0 & 0 & 0 
\end{array} \right).
\end{equation} 
 
\noindent The matrix can be abbreviated for convenience, by 
\begin{equation}\label{condensed}
J(\vec{\mu}, \vec{z}) = \left( \begin{array}{cc}
\hat{\mu} & \hat{z}\\
\hat{z} & 0
\end{array} \right),
\end{equation}
where the hatted entries correspond to the sub-blocks in the above matrix (or in other words, $\hat{\mu}$ could be viewed as the isomorphism between a cross product and its antisymmetric matrix representation in the semidirect algebra). 
 
The rank of the matrix~\eqref{sdpgravitymatrix} is easily seen to be 4. There are two Casimirs, $C_{1} = \vec{\mu} \cdot \vec{z}$ and $C_{2} = \|\vec{z}\|^{2}$ generically, which means that in Darboux' coordinates, the maximal symplectic leaves have dimension 4. When $z_{i} \equiv 0\mbox{ } \forall i$, then the matrix reduces to that of a free rigid body. A new subcasimir function that arises is $C_{3} = \|\vec{\mu} \|^{2}$. The leaves are 2-spheres, as in the rigid body case, with a totally degenerate origin when all subcasimirs collapse to arbitrary functions of $\mu$ and $z$ which intersect at zero.  
 
Consider an extension of the above to include one more force that operates on an underwater vehicle, namely buoyancy, so there are more rank changes here~\cite{marsdenpkcity}. Denote the buoyancy vector by $\vec{b} = (b_{1}, b_{2}, b_{3})$, the phase space vector by $(\vec{\mu}, \vec{z}, \vec{b})$, the semidirect product group by $W = SO(3) \ltimes (\Bbb{R}^{3}, \Bbb{R}^{3}) \simeq (SO(3) \ltimes \Bbb{R}^{3}) \ltimes \Bbb{R}^{3}$, with the group product operation defined as:
\begin{equation*}
(R, v, w) \cdot (R', v', w') = (RR', Rv'+v, Rw'+w).
\end{equation*}

The coadjoint space is given by $\frak{w}^{*} = \frak{so(3)} \ltimes \Bbb{R}^{3} \ltimes \Bbb{R}^{3}$, with the Lie-Poisson structure (in the condensed notation of~\eqref{condensed})
\begin{equation*}
J(\vec{\mu}, \vec{z}, \vec{b}) = \left( \begin{array}{ccc}
\hat{\mu} & \hat{z} & \hat{b}\\
\hat{z} & 0 & 0\\
\hat{b} & 0 & 0
\end{array} \right),
\end{equation*}
which is now a 9 by 9 matrix. At a generic point the coadjoint orbit has dimension 6, as there are 3 Casimirs $C_{1} = \vec{b} \cdot \vec{z}$, $C_{2} = \|\vec{z}\|^{2}$, $C_{3} = \|\vec{b}\|^{2}$. Nongeneric orbits come about in two stages of rank change:

\begin{enumerate}
\item 4-dimensional orbits that occur when $\vec{z}\mbox{ }||\mbox{ }\vec{b}$, which leads to two subcasimirs, $C_{4} = \vec{\mu} \cdot \vec{z}$, $C_{5} = \vec{\mu} \cdot \vec{b}$. 
 
\item 2-dimensional orbits (akin to free rigid body) that arise when $\vec{z} = \vec{b} = 0$, with the nontrivial subcasimir $C_{6} = \|\vec{\mu}\|^{2}$. As before, the origin in all 9 coordinates is a totally degenerate point. 
\end{enumerate}         
  
\subsection{Moment algebra for $(2+1)$-D Ideal Fluid}~\label{infinite}
 
This example has several rich and mysterious features which shall be explored in another publication~\cite{tomphilviv}. Here we confine the discussion to rank change. For the physics background, and details of the invariant theory computations, see~\cite{selfthesis}. 
 
Ideal fluids are infinite-dimensional, but they do have a Hamiltonian structure in function space~\cite{pjmrevu}. We consider the dynamics of the scalar {\em vorticity}, $\omega(x,y,t) = \hat{z} \cdot \nabla \ti v$, where $v$ is the Eulerian velocity, assumed to be divergence free, $\nabla \cdot v = 0$. In infinite dimensions, the Poisson bracket is defined in a functional setting. If $F, G \in C^{\infty}(\frak{B})$, with $\frak{B}$ some Banach space, are functionals of variables $\{\omega^{i}\}$, $i = 1, \ldots, n$, which are real-valued functions, then 
\begin{equation*}
\{F, G\} = \int_{D} \frac{{\delta}F}{{\delta \omega^{i}}} \frak{J}^{ij} \frac{{\delta}G}{{\delta \omega^{j}}} d \mu = \left \la \frac{{\delta}F}{{\delta \omega}},  \frak{J}  \frac{{\delta}G}{{\delta \omega^{j}}} \right \ra
\end{equation*} 
is the Poisson structure. It satisfies the Leibniz rule~\eqref{jac1} and Jacobi identities~\eqref{jac2}, besides --- owing to integration by parts --- being manifestly skew-symmetric. For the scalar vortex, there is a Poisson bracket (cf.~\cite{morrison81a, morrison82},~\cite{marswein83}) given by
\begin{equation}\label{vortexpb}
\{F, G\} = \int_{D} \omega \left[ \frac{{\delta}F}{{\delta \omega}}, \frac{{\delta}G}{{\delta \omega}} \right]dxdy = \left \la \omega, \left[ \frac{{\delta}F}{{\delta \omega}}, \frac{{\delta}G}{{\delta \omega}} \right] \right \ra,
\end{equation}
which has a form that resembles the Lie-Poisson bracket~\eqref{liepoisson}, with the inner Lie bracket being a function space tangent algebra bracket. Any arbitrary functional of the form $\int_{D} C(\omega) dxdy$ is a Casimir for the above bracket. One of the interesting facts about the scalar vortex is that the evolution dynamics can be studied entirely in terms of its projection down to a finite dimensional space of {\em moments}~\cite{kidamorrison}. This process is akin to the reduction in variables of a phase space due to the presence of a symmetry. (In this case the symmetry group is that of all volume preserving diffeomorphisms.) The reduced dynamics can be studied in its own right and projected back onto the original (infinite-dimensional) space to recover the full dynamics.

We suppress the integration measure and domain of integration in what follows. The moment description involves a space of moments, $(a_{1}, a_{2}, a_{3})$, where 
\begin{eqnarray}\label{amoments}
a_{1} & = & \int \frac{1}{2} x^{2} \omega, \notag \\
a_{2} & = & \int xy \omega, \notag \\
a_{3} & = & \int \frac{1}{2} y^{2} \omega.
\end{eqnarray}

\noindent The moments could easily be shown to satisfy the following commutation relations, where we use the Lie-Poisson structure~\eqref{vortexpb} to evaluate them:
\begin{eqnarray}\label{sl2algebra}
\{ a_{1}, a_{2} \} & = & 2a_{1} \notag, \\
\{ a_{2}, a_{3} \} & = & 2a_{3} \notag, \\
\{ a_{3}, a_{1} \} & = & - a_{2}, 
\end{eqnarray}
which is the standard commutation algebra of the Lie algebra of $SO(2,1) \simeq SL(2, \Bbb{R})$. The Casimir for this Lie algebra, with the metric induced by the Killing form of signature (1, 1, -1), is $C = a_{1}^{2} + a_{2}^{2} - a_{3}^{2}$. Its level sets are now in several components: a noncompact hyperboloid in one sheet for nonzero positive values of $C$, noncompact hyperboloids in two sheets for nonzero negative values of $C$, two components of a cone sans vertex for $C=0$ (but ${a_i}$ not all zero simultaneously), and the singular origin, for $C=0$ (and ${a_i}$ all zero simultaneously). The topology of the level sets is very different from that of a rigid body's Casimir levels, which are a union of compact 2-spheres and a singular origin. But we note that the {\em number} of Casimirs and the {\em dimension} of the generic leaves is the same for either case. 

In physics, when a field theory system is fully describable by a finite set of moments, it is said to have the property of {\em closure}. This moment algebra reduces the infinite dimensional scalar vortex dynamics to that of a three parameter vortex patch, known as a Kida vortex~\cite{kidamorrison, pjmrevu}. 
 
Now, let us consider adding a term proportional to the density, $\rho(x,y)$ to the Hamiltonian. This procedure is called {\em stratification}, in the event that the density varies in the vertical direction alone.  We consider a moment expansion for the density profile,
\begin{equation*}
\rho(x,y) \sim \sum b_{ij},
\end{equation*}
with $b_{ij}$ representing the moments for density:
\begin{equation}\label{bmoments}
b_{ij} = \int x^{i}y^{j} \rho,
\end{equation}   
where $i, j = 0,1, \ldots$. 

We note that $b_{ij}$ are essentially monomials in $x$ and $y$, so we can form a series of vector spaces, $\{\Bbb{V}^{k}\}$ consisting of all moments of homogeneous degree $k$. For example, $\Bbb{V}^{2} = (b_{20}, b_{11}, b_{02})$ for quadratic order. In general, $\Bbb{V}^{k}$ has $k+1$ monomials.  
 
The algebra of homogeneous polynomials under the Poisson bracket like~\eqref{vortexpb} would display the following derivation rule,
\begin{equation*}
\{\mbox{ },\mbox{ } \}: \Bbb{V}^{m} \ti \Bbb{V}^{n} \lo \Bbb{V}^{(n+m-2)}
\end{equation*}
so that only the tensor product of all these spaces $\Bbb{V}^{1} \otimes \Bbb{V}^{2} \otimes \cdots$ is closed under the bracket (except for $n=m=3$). The situation however changes when we {\em define} the following semidirect generalization of the Lie-Poisson bracket~\eqref{vortexpb}     
\begin{equation}\label{vortexsdp}
\{ F, G \} = \int \omega[F_{\omega}, G_{\omega}] + \rho([F_{\omega}, G_{\rho}] - [G_{\omega}, F_{\rho}]),
\end{equation}
with $F, G$ arbitrary functionals of both the vorticity and the density, and the subscripts on them with respect to these arguments stands for the operation of functional derivative. 
 
It turns out that~\cite{tomphilviv} the moment algebra of the stratified fluid with the above semidirect algebra, is actually closed to any order of truncation of the density moments~\eqref{bmoments}. That is, the space of monomials given by $\{a_{1}, a_{2}, a_{3}\} \cup \{ \Bbb{V}^{k} \}$ for any $k$, is closed with respect to the above bracket. Hence, the dynamics of the stratified fluid may be projected down to its space of moments order by order, to any desired level of approximation.

We note that the above bracket~\eqref{vortexsdp} represents the {\em dual} to the semidirect product of $\frak{sl(2)}$ with $\Bbb{V}^{k}$ via the adjoint action. That is, we collect together the moment algebra and take their Poisson brackets mutually to get a structure matrix representing the semidirect product of the dual algebra. For the case of $k =1$, we would get a 5 by 5 matrix of quadratic $a_{i}$ moments, and linear $b_{01}$, $b_{10}$ moments. There is one Casimir. 

For the case of $k = 2$, we would end up with a 6 by 6 matrix that is virtually identical to the semidirect product of the rigid body in gravity~\eqref{sdpgravitymatrix} (except for scaling, signature, and a linear transformation to diagonal coordinates). There are two quadratic Casimirs, just as for~\eqref{sdpgravitymatrix} (except with one minus sign in each), so similar comments regarding rank change and subcasimirs apply. The Casimirs are $C_{6}^{1} = b_{11}^{2} - b_{20}b_{02}$ and $C_{6}^{2} = a_{1}b_{02} - a_{2}b_{11} + a_{3}b_{20}$.

The Casimirs at any given moment truncation will bring the system down to a certain 6 dimensional manifold. For instance, the Casimir for $J_{7}$ on $\bv^{7}$ is a quartic, given by $C_{7} = 4b_{12}^{3}b_{30} + 4b_{21}^{3}b_{03} - 6b_{03}b_{12}b_{21}b_{30} + b_{03}^{2}b_{30}^{2} -3b_{12}^{2}b_{21}^{2}.$ The two Casimirs of $J_{8}$ are a quadratic and a cubic, given by $C_{8}^{1} = b_{04}b_{40} - 4b_{13}b_{31} + 3b_{22}^{2}$ and $C_{8}^{2} = -b_{22}^{3} + 2b_{13}b_{22}b_{31} - b_{04}b_{31}^{2} - b_{13}^{2}b_{40} + b_{04}b_{22}b_{40}$.

Finally, it is worth emphasizing that all these various Casimirs and the subcasimirs can be determined using invariant theory computations alone, following~\cite{mihailovs}, without so much as referring to the Poisson structure. The various computer algebra packages do not yield compact expressions for the Casimirs as above without some effort in finding integrating factors by guesswork~\cite{selfthesis}.  

\section{Conclusion}\label{conclusion}

The problem of rank change in Poisson dynamical systems is important for two reasons. Systems tend to prefer equilibrium states that are more constrained (by the available degrees of freedom). Also, a full examination of stability of equilibria requires an assessment of subcasimirs that arise upon rank change. Ideas from orbit geometry and invariant theory are invoked, with the underlying assumption that systems with some sort of symmetry and invariance properties prove physically and geometrically interesting. Casimirs and subcasimirs can be found either by examining the Poisson structure directly, or by using invariant theory methods, as we show in some examples considered.

Quite apart from all this, symmetry breaking ideas are introduced. A potential invariant under the symmetry group is equipped with control parameters. The new Goldstone modes which appear upon changing the orbit type (using the controls), are shown to be the same extra data provided by subcasimirs to pin down singular orbits. This demonstration is made through the geometry of strata in orbit space. Some classes of orbits are seen to carry the same information as the Poisson structure. In this way, using the glue of orbit geometry and the associated invariant theory, we see the overlap between two seemingly different descriptions of the same geometric object. 

It would be nice to do away with the compactness requirement of the symmetry group, which we needed to keep track of the dimensions of the invariant normal directions in the orthogonal decomposition in Section~\ref{tangnorm}. From simply counting the number of Casimirs and subcasimirs, all the examples seem to go through for both the compact and noncompact groups. Even the strange phenomena reported in the moment truncation example, Section~\ref{infinite}, work analogously for the cases of $SO(3)$ and $SO(2,1)$ (see~\cite{tomphilviv} for details). This is strong evidence that compactness plays no role in the identification of kinematical degrees of freedom we propose in this article. The actual dynamics, as specified by the evolution of Hamilton's equations would of course differ in the noncompact case, for instance, Poincare recurrence will no longer apply. 

Another issue for future work is using the mass matrix stability to address some of the stability issues usually addressed using the Energy-Casimir procedure. The mass matrix does not truly depend on a potential, for all $G$-invariant potentials sample the same orbit geometry of whose invariants they are constituted, so by an appropriate choice of potential, various facets of the orbit geometry, hence possible Poisson structures, may be studied.    

\section*{Acknowledgments}
This research was supported by the US Department of
Energy Contract No.~DE-FG03-96ER-54346.


\end{document}